\documentclass[amsmath,amssymb,aps,letterpaper,pre,twocolumn,longbibliography,superscriptaddress]{revtex4-2}
\usepackage{tikz}
\usepackage[english]{babel}
\usepackage[utf8]{inputenc}
\usepackage[T1]{fontenc}
\usepackage{indentfirst}
\usepackage{amsmath}
\usepackage{amssymb}
\usepackage{amsthm}
\usepackage{proof}
\usepackage{eufrak}
\usepackage{graphicx}
\usepackage{psfrag}

\allowhyphens

\newcommand{\km}{{\boldsymbol k}_M}

\newcommand{\pn}{{\boldsymbol p}_N}

\newcommand{\dipo}{\mathcal{D}}
\newcommand{\UU}{\mathcal{U}}
\newcommand{\La}{\mathcal{L}}
\newcommand{\Pe}{\mathcal{P}}
\newcommand{\complex}{\mathbb{C}}

\newcommand{\ket}[1]{{\left|#1\right\rangle}}


\usepackage{ifpdf}

\ifpdf
\usepackage{epstopdf}
\usepackage[pdftex,colorlinks,urlcolor=blue,citecolor=blue,linkcolor=blue]{hyperref}
\else
\usepackage[hypertex,colorlinks,urlcolor=blue,citecolor=blue,linkcolor=blue]{hyperref}
\fi
\begin{document}
\numberwithin{equation}{section}

\title{Integrable hard rod deformation of the Heisenberg spin chains}
\author{Bal\'azs Pozsgay}
\affiliation{MTA-ELTE “Momentum” Integrable Quantum Dynamics Research Group, Department of Theoretical Physics, Eötvös Loránd University}
\author{Tam\'as Gombor}
\affiliation{MTA-ELTE “Momentum” Integrable Quantum Dynamics Research Group, Department of Theoretical Physics, Eötvös Loránd University}
\affiliation{Holographic QFT Group, Wigner Research Centre for Physics, Budapest, Hungary}
\author{Arthur Hutsalyuk}
\affiliation{MTA-ELTE “Momentum” Integrable Quantum Dynamics Research Group, Department of Theoretical Physics, Eötvös Loránd University}

\begin{abstract}
We present new integrable models of interacting spin-1/2 chains, which can be interpreted as hard rod deformations of
the XXZ Heisenberg chains. 
The models  support multiple particle types: dynamical hard rods of length $\ell$ and
particles with lengths $\ell'<\ell$ that are immobile except for the interaction with the hard rods. We encounter a
remarkable phenomenon in these interacting models: exact spectral degeneracies across different deformations and
volumes.
The algebraic integrability of these systems is also
treated using a recently developed formalism for medium range integrable spin chains. We present the detailed Bethe Ansatz
solution for the case $\ell=2$.
\end{abstract}

\maketitle

\section{Introduction}

The $T\bar T$ deformation of Quantum Field Theories \cite{Zam-TTbar,Smirnov-Zam-TTbar,Cavaglia:2016oda,zam-genttbar} together with
certain generalized deformations have attracted
considerable interest in the last couple of years, see the review paper \cite{Jiang:2019hxb}.
In Field Theory the generalized $T\bar T$ deformations can be formulated for any pair of local conserved
charges, and for integrable Field Theories this gives new possibilities for integrable deformations. 
Very recently it has been argued that this enlarged family of
deformations is so rich that it spans the tangent space of integrable systems \cite{doyon-space}. In other
words, if there 
are two integrable models that can be transformed into each other continuously, then it should be possible to perform
this transformation using only the generalized $T \bar T$ deformations.

There are various geometrical interpretations of the $T\bar T$ deformation, and one of them is that
the fundamental particles acquire a momentum dependent width
\cite{ttbar-geom1,ttbar-geom2,doyon-cardy-ttbar,rule54-ttbar,yunfeng-hardrod-1,yunfeng-hardrod-2}. Similar
interpretation can be given also for the generalized deformations. A particular case is the so-called hard rod
deformation, where the particles acquire a finite and fixed width. This implies that starting from a free theory one can
build quantum mechanical or field theoretical models  of moving hard rods \cite{yunfeng-hardrod-1,yunfeng-hardrod-2}.

The study of the hard rod systems has a long history, going back to the classical physical hard rod gas 
\cite{nagamiya-hardrod,rubin-hardrod,sutherland-hardrod}. This is a many body system with a very simple 
interaction: 
the hard rods move freely as long as they do not collide. The collisions are elastic, 
and if the rods are indistinguishable then a collision can be interpreted as a
finite displacement of the particle coordinates given directly by the hard rod length. 
This simplicity of the interaction lead to the proofs that hydrodynamical behaviour emerges in the thermodynamic limit
  \cite{hard-rod-gas,hard-rod-gas2}, and this is a very important direct link between the microscopic
  and macroscopic descriptions of many body systems (see also \cite{doyon-spohn-hardrod}). This success motivates the
search for analogous models in the 
  quantum realm, in the hope that emergent hydrodynamics could be established rigorously, despite having non-trivial
  interactions in the system.

An obvious question one can ask is whether the $T\bar T$ deformation and the various generalizations can be
formulated 
for lattice systems
or non-relativistic 
field theories.
In \cite{sajat-ttbar,sfondrini-ttbar} it was found that the generalized $T\bar T$ deformations do exist on integrable
spin chains. What is more, 
they are special cases of a large family of long range deformations that were studied in the context of the
AdS/CFT correspondence  a decade earlier \cite{beisert-long-range-2}. However, the results
\cite{sajat-ttbar,sfondrini-ttbar} can not be applied for those transformations which would involve the momentum
operator, simply because there is no local momentum operator on the lattice.
This means in particular that the actual $T\bar T$ deformation and the hard rod deformation (which involve the momentum)
are not included in the
possibilities provided by the formalism of \cite{sajat-ttbar,sfondrini-ttbar}. This motivates further research towards
the lattice realization of these deformations.

In this paper we investigate the hard rod deformation for lattice systems, and we show that the deformation can be
performed for the  
integrable XXZ spin chains, such that it preserves the integrability of the models.

We pose two main requirements for the hard rod deformation: 
a) the fundamental
particles should acquire a finite width, and b) the Hilbert space of the new model should have the same structure as that
of the original model, i.e. it should be a simple tensor product space. The second one is a very important
requirement. Earlier works 
already studied a number of models where the particles show hard rod behaviour \cite{constrained1,constrained2,constrained3,constrained5,constrained-multi,bariev-alcaraz-hubbard-tj0,bariev-alcaraz--hubbard-tj,alcaraz-bariev-second-class,alcaraz-bariev-second-class2,hardcorebariev1,susy-constrained-lattice-hopping1,susy-constrained-lattice-hopping2}, but there the Hilbert space was restricted so
that the kinetic constraints coming from the finite length are automatically satisfied.
Such models can be interesting for a variety of reasons, see for example the recent works \cite{borla--constrained,pollmann-bond-site-xxz,alvise-constrained}.
However, we intend to find
local Hamiltonians such that the hard rod behaviour emerges dynamically, and it is not encoded in the Hilbert space
structure. Naturally this means that if the dynamical hard rods have length $\ell$ then particles with lengths $\ell'<\ell$ should
also be present in the system, otherwise we could not produce a complete Hilbert space of the tensor product form. 

A particular model with the desired requirements is the ``folded XXZ model'' which was studied in the recent works
\cite{folded1,folded2,sajat-folded,folded-jammed}. It was explained in \cite{sajat-folded} that it can be regarded as
the hard rod deformation of the XX chain. The folded XXZ model has a number of interesting dynamical features, such as
exponentially large degeneracies for excited states, Hilbert space fragmentation, particle-hole transmutation
\cite{quantum-bowling} and persistent oscillations after certain quantum quenches; its hydrodynamic behaviour was
treated in \cite{folded2}.
It is our goal in the present paper
to give a generalization of the ``folded XXZ model'' such that it becomes the hard rod deformation of the interacting
XXZ models. 

To this order we used the recent methods of \cite{sajat-medium} where a general framework was given for spin chains with
medium range interactions. In fact, the Hamiltonian of the hard rod deformed XXZ model with rod length $\ell=2$ was
already announced in
that work. Here we analyze this model in detail and also give the hard rod deformation for arbitrary rod length $\ell\ge
2$. 

In the next Section we summarize our main results, leaving the technical details to the later Sections. The hard rod
deformation with $\ell=2$ is discussed first in Section \ref{sec:ell2}, the coordinate Bethe Ansatz solution is given
with open and periodic boundary conditions in \ref{sec:cool2}, and the algebraic construction ensuring its integrability
is presented in Section \ref{sec:aba}. In this Section we also treat the so-called chiral charges of the folded XXZ
model, for which there was no algebraic result available in the previous literature.
The hard rod deformed model with $\ell=3$ is introduced and briefly discussed in
\ref{sec:ell3}, and Section \ref{sec:concl} includes our conclusions.

\section{Summary of the main results}

\label{sec:main}

In this Section we summarize some key properties of the original XXZ spin chain, and we also present the main results
for the hard rod deformations. The solution of the new models and the discussion of its algebraic properties are given
in later Sections.

\subsection{Notations}

We treat spin-1/2 chains. Let $\complex^2$ be the local Hilbert space at a
site of the chain, and let $\ket{\uparrow}$ and $\ket{\downarrow}$ be the usual basis vectors. Furthermore
$\sigma^{x,y,z}$ denote the Pauli matrices, and $\sigma^{\pm}$ are the raising and lowering operators. The models that
we treat conserve the global $S^z$ quantum numbers, but they are not symmetric with respect to a global spin flip,
therefore we need to make a clear distinction between the up and down spins. We interpret the up spins as the vacuum and
the down spins as particles, and throughout this work we use the following alternative notation for the basis vectors:
\begin{equation}
  \ket{\circ}\equiv\ket{\uparrow},\qquad \ket{\bullet}\equiv\ket{\downarrow}.
\end{equation}
Furthermore we introduce the local projectors onto these states:
\begin{equation}
  P^\circ=\frac{1+\sigma^z}{2},\qquad   P^\bullet=\frac{1-\sigma^z}{2}.
\end{equation}
The reference state is defined as $\ket{\emptyset}=\ket{\circ\circ\circ\dots}$.

We also define a particle number operator as
\begin{equation}
\mathcal{N}=\sum_j P^\bullet_j.
\end{equation}
All our Hamiltonians preserve $\mathcal{N}$.

\subsection{The Heisenberg chain}

The XXZ Heisenberg spin chain is defined as
\begin{equation}
  H=\sum_j \left[ \sigma^x_j\sigma^x_{j+1}+ \sigma^y_j\sigma^y_{j+1}+\Delta (\sigma^z_j\sigma^z_{j+1}-1)\right].
\end{equation}
We chose an additive normalization such that the reference state is an eigenstate with zero energy. The real number
$\Delta$ is the so-called anisotropy parameter.

Using the operators
introduced above an alternative representation of the XXZ Hamiltonian is
\begin{equation}
  H=2\sum_j h_{j,j+1}
\end{equation}
with
\begin{equation}
  \label{XXZh}
  h_{j,j+1}=\sigma^-_j\sigma^+_{j+1}+\sigma^+_j\sigma^-_{j+1}
    -\Delta (P^\circ_j P^\bullet_{j+1}+P^\bullet_j P^\circ_{j+1}).
\end{equation}

The interpretation of this Hamiltonian is straightforward: The first two terms generate propagation of particles to the
left and to the right, and the third term describes the interaction between particles. It is useful to discuss how the
interaction emerges. The interaction term is diagonal in the computational basis and it gives a contribution of
$-\Delta$ for each pair of neighbouring sites if one of 
them is empty and one 
of them is occupied. For a single particle state the summation over these diagonal terms gives
$-2\Delta$. For two particle states the diagonal contribution is $-4\Delta$ if the particles are at distance more than
1, and $-2\Delta$ if they occupy neighbouring positions. This difference is the source of the interaction between them.

The XXZ spin chain is integrable and is possesses a set of local conserved charges $\{Q_\alpha\}$ which are indexed by
an integer $\alpha\ge 1$. The charges are extensive and can be expressed as
\begin{equation}
  Q_\alpha=\sum_j q_{\alpha}(j),
\end{equation}
where $q_{\alpha}(j)$ is the operator density positioned at site $j$. We choose a normalization such that the range of
the $q_{\alpha}(j)$ is $\alpha$ and it spans the sites $j\dots j+\alpha-1$. Thus $Q_1$ can be identified with the particle
number operator $\mathcal{N}$, and $Q_2$ is 
identified with (or proportional to) the Hamiltonian.  The charges form a commuting set:
\begin{equation}
  [Q_\alpha,Q_\beta]=0.
\end{equation}
They can be constructed either using the so-called boost operator \cite{Tetelman,Thacker-boost,GM-higher-conserved-XXZ},
or from a commuting set of transfer matrices. Below we will review the latter construction, which is based on the famous
Yang-Baxter relation.

It is an interesting consequence of the Yang-Baxter integrability, that (up to
normalization) the density of the next conserved charge $Q_3$ can be expressed as
\begin{equation}
  \label{q3relation}
  q_{3}(j)=[h(j),h(j+1)],
\end{equation}
where we used the simplified notation $h(j)=h_{j,j+1}$.
This relation seems to hold in every nearest neighbour integrable model
\cite{kulish-sklyanin-developments,Kennedy-all-su2-chains,Mutter-Schmidt-classification,integrability-test,marius-classification-1}.  

\subsection{Hard rod deformed models}

We introduce a new integrable hard rod deformation of the XXZ Heisenberg spin chains. The idea is that the main
propagating degrees of freedom should be ``hard rods'' of length $\ell$, where a hard rod is simply a sequence of
$\bullet$ states, with $\ell\ge 1$. 

Our requirements for the
construction of the model are the following:
\begin{enumerate}
\item The model should describe the propagation of the hard rods of length $\ell$, which should be stable particles.
\item The Hilbert space of the model should be the standard tensor product space of the spin chains.
\item The Hamiltonian should be strictly local: it should have a Hamiltonian density which is a short range
  operator, but the range can depend on the hard rod length $\ell$.
\item The model should be integrable: it should have a set of conserved charges and it should be solvable by the Bethe Ansatz.
\end{enumerate}

We propose the following family of Hamiltonians with integer parameter $\ell$, which satisfy all requirements:
\begin{equation}
  \label{Hhrdefgen1}
  H=\sum_j  \left[ h_{j,j+\ell} \prod_{k=1}^{\ell-1} P^\bullet_{j+k}\right].
\end{equation}
Here the two-site operator $h_{j,j+\ell}$ is the same as in \eqref{XXZh} but now it acts on sites that are at distance
$\ell$ from each other.  The insertion of the projection operators between sites $j$ and $j+\ell$ can
be understood as a control for the action of the original Hamiltonian density $h_{j,j+\ell}$. The case of $\ell=1$
corresponds to the original XXZ chain. 

Substituting \eqref{XXZh} into \eqref{Hhrdefgen1} we obtain two terms: a kinematical and an interaction term.

The kinematical term in \eqref{Hhrdefgen1} generates the following moves (transition matrix elements) on a segment of $\ell+1$ sites:
 \begin{equation}
   \underbrace{\bullet\cdots\bullet}_{\ell }\circ\qquad \leftrightarrow\qquad
\circ    \underbrace{\bullet\cdots\bullet}_{\ell }.
\end{equation}
We can see that this term moves the hard rods of length $\ell$ on the chain, but it leaves sequences of $\bullet$ states
with length shorter 
than $\ell$ invariant. This means that particles of length $\ell'<\ell$ are not dynamical on their own. However, they
can become dynamical through their interaction with the hard rods of length $\ell$.

The interaction term is
\begin{equation}
  -\Delta  \sum_j (P^\circ_j P^\bullet_{j+\ell}+P^\bullet_j P^\circ_{j+\ell}) \left(\prod_{k=1}^{\ell-1} P^\bullet_{j+k}\right).
\end{equation}
In the computational basis this operator measures the number of separate segments consisting only of $\bullet$
states with length greater than $\ell$. Thus it gives an interaction term for hard rods of length $\ell$, similar to how
the interaction works in the nearest neighbour case. On the other hand, this operator does not take into account shorter
immobile particles with length $\ell'<\ell$.

\subsection{Connection with other models}

Let us now discuss the connections with some models that appeared earlier in the literature.

The idea of describing particles with finite width on a spin-1/2 chain appeared much earlier in the works
\cite{constrained1,constrained2,constrained3}, which treated the so-called constrained XXZ models. Here the idea is to
have just a single down spin (or $\bullet$) as the fundamental particle, but the models are engineered to mimic a finite
width. This is achieved by forbidding that two particles occupy sites that are closer to each other than the given
minimum length $\ell$. The Hamiltonian is then written as
\begin{equation}
  \label{constrainedH}
  H=P^{(\ell)} \left[\sum_j \sigma^x_j\sigma^x_{j+1}+\sigma^y_j\sigma^y_{j+1}+
\Delta    \sigma^z_j\sigma^z_{j+\ell} \right] P^{(\ell)},
\end{equation}
where $P^{(\ell)}$ is a global projector that selects those states in the Hilbert space where no two down spins are
closer to each other than $\ell$. In the case of $\ell=1$ this projector is just the identity, and we obtain simply
the XXZ Hamiltonian. For $\ell\ge 2$ we obtain a new solvable model, whose integrability was established both by
coordinate Bethe Ansatz, and by a long range vertex model \cite{constrained3}.

The relation between \eqref{constrainedH} and our  models is the following. In our model the hard rods are represented by
a sequence of $\bullet$'s, but sequences with arbitrary lengths are allowed and there is no global projector. However,
if we project our model to the sector that consists only of the hard rods with length $\ell$ and no other lengths, then
we obtain a Hamiltonian identical to \eqref{constrainedH}. This is seen by directly computing the matrix elements of the
two Hamiltonians. 

We believe that one of the advantages of our new models is that the Hamiltonians are completely local and there is no
need for a global projector to define the theory.
For other closely related constrained models see for example
\cite{constrained5,constrained-multi,bariev-alcaraz-hubbard-tj0,bariev-alcaraz--hubbard-tj,alcaraz-bariev-second-class,alcaraz-bariev-second-class2,hardcorebariev1}
and the super-symmetric spin chains treated in
\cite{susy-constrained-lattice-hopping1,susy-constrained-lattice-hopping2}.

\subsection{Conserved charges}

The hard rod deformed models belong to the family of ``medium range'' spin chains treated in detail in
\cite{sajat-medium}. Therefore the general algebraic methods developed in \cite{sajat-medium} can be applied in these
cases as well.

Using these methods we found that the models given by \eqref{Hhrdefgen1} also possess a family of conserved charges
$\{Q_\alpha\}_{\alpha \in \mathcal{S}}$ where 
now $\mathcal{S}$ is a subset of the integers.  The particle number operator is always conserved, thus $1\in
\mathcal{S}$. The next charge is the Hamiltonian, which spans $\ell+1$ sites, thus we identify
\begin{equation}
  Q_{\ell+1}=H.
\end{equation}
The next few charges have range $2\ell+1,3\ell+1,\dots$, thus the allowed set of  charge
indices for a given hard rod length $\ell$ is
\begin{equation}
\mathcal{S}=\{1,\ell+1,2\ell+1,3\ell+1,\dots\}.
\end{equation}
The situation that different models can have charges with different indices is very well known from integrable QFT,
where the indices correspond to the ``spin'' of the charge under Lorentz boost \cite{Mussardo-review}. 

\subsection{Special dynamical properties}

Below we show that these hard rod deformed models can be solved by coordinate Bethe Ansatz, both with periodic and with
open boundary conditions. The latter case will prove to be simpler.

We observe an interesting phenomenon: {\it spectral
  degeneracies across different deformations and different volumes}. This means that the spectrum of a hard rod deformed
model with parameters $\Delta$ and 
$\ell$ in a volume $L$  consists of the spectra of the original XXZ chain with the same $\Delta$ and various volumes
$L'\le L$. And while the original XXZ chain does not have degeneracies on top of those required by symmetry, the energy
levels of the hard rod deformed models are typically exponentially degenerate with respect to the volume, with an
exponent that depends on the physical content of the excited state in question.

The phenomenon of spectral
  degeneracies across different deformations and different volumes might seem natural from the point of view of classical physics, where the only effect of the
hard rod length is the modification of the volume available to the particles. However,
it is remarkable that the degeneracies hold exactly even in the quantum case. All our Hamiltonians
 are local, and we do not 
exclude particles with length
$\ell'<\ell$, therefore it is indeed remarkable to find exact equalities between the energy levels. We stress in advance
that these spectral degeneracies connecting different deformations and different volumes hold only if open boundary
conditions are applied; in  
the periodic case they only hold for a subset of energy levels.

The exponential degeneracy of a given level of a deformed model is caused by the various ways of how we can add particles
with length $\ell'<\ell$ to a given configuration of dynamical hard rods. This can be interpreted as {\it Hilbert space
  fragmentation}: the presence of the particles with length $\ell'<\ell$ fragments 
the Hilbert space into disconnected sectors.  Typically the 
degeneracies grow exponentially with the volume, but the exponent depends on the particle content. The mechanism for the
degeneracies is the same as discussed in \cite{folded1,folded2,sajat-folded}, and it is discussed also in Section \ref{sec:cool2}.

We expect that our new models also display persistent oscillations after specific quantum quenches, similar to the
folded XXZ model \cite{sajat-folded}. However, we do not treat quench problems in this work.

\section{Hard rod deformation with $\ell=2$}

\label{sec:ell2}

Let us now focus on the hard rod deformed model with $\ell=2$.
In this case the Hamiltonian is
\begin{multline}
  \label{Hhrdef2}
   H=\sum_j  \left[\sigma^-_jP^\bullet_{j+1}\sigma^+_{j+2}+\sigma^+_jP^\bullet_{j+1}\sigma^-_{j+2}\right.\\
\left.    -\Delta (P^\circ_j P^\bullet_{j+1}P^\bullet_{j+2}+P^\bullet_j P^\bullet_{j+1}P^\circ_{j+2})\right].
\end{multline}
This model was announced in \cite{sajat-medium} as one of the $U(1)$-invariant three site interacting families of
integrable spin chains.

\subsection{Connection with existing models}

At the special point $\Delta=0$ we obtain a relatively simple Hamiltonian:
\begin{equation}
  \label{Hbond}
   H=\sum_j  \left[\sigma^-_jP^\bullet_{j+1}\sigma^+_{j+2}+\sigma^+_jP^\bullet_{j+1}\sigma^-_{j+2}\right].
\end{equation}
This coincides with a specific version of
the ``folded XXZ'' model treated recently in 
\cite{fracton1,folded1,folded2,sajat-folded}. To be precise, the Hamiltonian given by \eqref{Hbond} is the ``dual
model'' of \cite{folded1,folded2} and the ``bond picture'' model of \cite{sajat-folded}. It is useful to discuss this
connection in more detail.

The original ``folded XXZ model'' is defined by the four-site Hamiltonian \cite{maurizio-coreprere,folded1,folded2}
\begin{equation}
  \label{Q4folded}
  H=\sum_{j=1}^L \frac{1+\sigma^z_j\sigma^z_{j+3}}{2}( \sigma^+_{j+1}\sigma^-_{j+2}+ \sigma^-_{j+1}\sigma^+_{j+2}).
\end{equation}
This model is spin-flip invariant; its dynamics was discussed at length in \cite{folded1,folded2,sajat-folded}.
The Hamiltonian density
generates propagation 
of single particles between sites $j+1$ and $j+2$, such that the hopping amplitude is controlled by the state of the
sites $j$ and $j+3$. It was explained in \cite{sajat-folded}, that in this model the propagating particles are either a
single down spin embedded into a vacuum of up spins, or vice versa, a single up spin embedded into a sea of down
spins. Furthermore, one observes particle-hole transmutation \cite{quantum-bowling}.

The dynamics of \eqref{Q4folded} appears different from the one generated by \eqref{Hbond}, however the two models are
connected by a non-local transformation which we now describe.
The main idea is to perform a bond-site
transformation: we construct a non-local mapping from the Hilbert space of the model \eqref{Q4folded} to
that of \eqref{Hhrdef2} where we put variables on the bonds (links) between the spin chain sites. The mapping is worked
out in the computational basis. Taking a specific basis state, for each bond we write
down a $\bullet$ if the neighbouring sites have different spin, and a $\circ$ if they have identical spin. Explicit
computations with local basis states show that the Hamiltonian \eqref{Q4folded} is mapped to the three-site Hamiltonian
\eqref{Hbond}; this was performed in \cite{sajat-folded}. For other examples of this bond-site transformation see \cite{sajat-medium}.

It was already argued in \cite{sajat-folded} that the model given by \eqref{Q4folded}-\eqref{Hbond} should be regarded
as the hard rod deformation of the XX model, which is defined by the Hamiltonian
\begin{equation}
  \label{HXX}
   H_{XX}=\sum_j  \left[\sigma^-_j\sigma^+_{j+1}+\sigma^+_j\sigma^-_{j+1}\right].
\end{equation}
The hard rod deformation appears through the insertion of the projectors $P^\bullet$ between the hopping operators
in \eqref{HXX}, leading to \eqref{Hbond}.
The new addition of \cite{sajat-medium} and the present work is that the hard rod deformation can be extended to the
interaction terms as well, 
which gives the interacting Hamiltonian \eqref{Hhrdef2} starting from the original XXZ model given by \eqref{XXZh}.

It is useful to perform the bond-site
transformation backwards from \eqref{Hhrdef2}. Then we obtain the four site Hamiltonian
\begin{multline}
  \label{eq:4siteXXZ}
  H=\sum_j \frac{1+\sigma^z_{j}\sigma^z_{j+4}}{4}\times \\
  \times\left[
    \sigma^x_{j+1}\sigma^x_{j+2}+\sigma^y_{j+1}\sigma^y_{j+2}  +\Delta (\sigma^z_{j+1}\sigma^z_{j+2}-1)
  \right].
\end{multline}

Quite interestingly a similar Hamiltonian was also proposed in \cite{fracton1}. That work studied the model given by
\begin{multline}
  \label{eqfr}
  H=\frac{1}{8}\sum_j (1+\sigma^z_{j}\sigma^z_{j+4}) \left[
    \sigma^x_{j+1}\sigma^x_{j+2}+\sigma^y_{j+1}\sigma^y_{j+2}\right]+\\
  +\Delta \sum_j \sigma^z_{j}\sigma^z_{j+2}.
\end{multline}
We modified the normalization given in \cite{fracton1} to match our conventions. It was found in \cite{fracton1} that
the Hamiltonian \eqref{eqfr} is not integrable, although it has an integrable sector. This sector consists of those states where the
particle do not occupy neighbouring sites, and the sector is completely identical to the constrained XXZ model
defined by \eqref{constrainedH} with $\ell=2$.

Our results give an explanation for the non-integrable sectors of
\eqref{eqfr}: the interaction term added in \eqref{eqfr} is not compatible with integrability, and the only integrable term
is the one given in \eqref{Hhrdef2}.

It is worthwhile to mention one more connection with existing models in the literature. The so-called Bariev model
\cite{bariev-model} describes two coupled XX chains, which can be considered as a zig-zag spin ladder. Alternatively the
Hamiltonian can be written in a translationally invariant form with a three site interaction:
\begin{equation}
  \label{bariev}
  H=\sum_j
   \left[\sigma^-_j\sigma^+_{j+2}+\sigma^+_j\sigma^-_{j+2}\right] \frac{1-U\sigma^z_{j+1}}{2},
\end{equation}
where $U$ is a coupling constant (our multiplicative normalization differs from that of \cite{bariev-model} in the
factor of 1/2 that we added). We can see that the Bariev model becomes identical with the hard rod deformed XX model
\eqref{Hbond} at the special point $U=1$. This connection was already noted in \cite{folded1,folded2}.

The one-parameter families
of Hamiltonians \eqref{Hhrdef2} and \eqref{bariev} overlap only at one point, corresponding to $\Delta=0$ and $U=1$.

\subsection{Conserved charges}

The hard rod deformed model possesses an infinite set of conserved charges, which can be obtained from the algebraic
construction 
discussed in Sec. \ref{sec:aba}. Here we just give some simple remarks.

As discussed above, the global $S^z$ is conserved, therefore we identify the first charge as $Q_1=\mathcal{N}$.
Furthermore
we identify 
$Q_3=H$. Writing $H=\sum_j h_{j,j+1,j+2}$ we obtain the next non-trivial charges as
\begin{equation}
  Q_5=\sum_j  q_{5}(j)
\end{equation}
with
\begin{equation}
  \label{q5relation}
  q_{5}(j)=[h(j),h(j+1)+h(j+2)],
\end{equation}
where we used $h(j)=h_{j,j+1,j+2}$. 
This relation is a generalization of \eqref{q3relation} and it was derived in \cite{sajat-medium}.

For the concrete model we find
\begin{equation}
  q_{5,j}=i(q_{5,j}^--q_{5,j}^+),\qquad
  q_{5,j}^+=\left(q_{5,j}^-\right)^\dagger
\end{equation}
with
\begin{multline}
    q_{5,j}^-=
     \sigma_{j}^-\sigma^-_{j+1}\sigma^+_{j+2}\sigma^+_{j+3}+
     \sigma_j^- P^\bullet_{j+1}\sigma^z_{j+2}P^\bullet_{j+3}\sigma_{j+4}^++\\
    +2\Delta(-\sigma^-_j P^\bullet_{j+1}\sigma^+_{j+2}+P^\bullet_j P^\bullet_{j+1} \sigma^-_{j+2} P^\bullet_{j+3}\sigma^+_{j+4}+\\
+  \sigma^-_{j} P^\bullet_{j+1}\sigma^+_{j+2}    P^\bullet_{j+3} P^\bullet_{j+4} ).
\end{multline}
The commutativity of $Q_3$ and $Q_5$ can be checked by direct computation; we also used computer programs for 
checks of the formulas. 

In the special case of the hard rod deformed XX model (folded XXZ model) we find that there are actually two families of
charges that are conserved. We denote them by $Q^{\pm}_{\alpha}$, where $\alpha=3,5,7,\dots$.
These charges are ``chiral'', which means that they move hard rods only in one direction. This is discussed in
detail in Subsection \ref{sec:chiral}.

\section{Coordinate Bethe Ansatz for $\ell=2$}

\label{sec:cool2}

In this Section we present the coordinate Bethe Ansatz solution of the model. The solution is a simple generalization of
the one presented in \cite{sajat-folded} for the hard rod deformed XX model. 

\subsection{Coordinate Bethe Ansatz -- open boundary conditions}

\label{sec:boundary}

We consider the XXZ and the deformed XXZ models with free boundary conditions. The Hamiltonians are defined now as
\begin{align}
H_{L}^{(1)}
  =\sum_{j=1}^{L-1}\sigma_{j}^{+}\sigma_{j+1}^{-}+\sigma_{j}^{-}\sigma_{j+1}^{+}-\Delta\left(P_{j}^{\circ}P_{j+1}^{\bullet}+P_{j}^{\circ}P_{j+1}^{\bullet}\right)
\end{align}
and
\begin{multline}
H_{L}^{(2)} 
              =\sum_{j=1}^{L-2}\sigma_{j}^{+}P_{j+1}^{\bullet}\sigma_{j+2}^{-}+\sigma_{j}^{-}P_{j+1}^{\bullet}\sigma_{j+2}^{+}-\\
-  \Delta\left(P_{j}^{\circ}P_{j+1}^{\bullet}P_{j+2}^{\bullet}+P_{j}^{\bullet}P_{j+1}^{\bullet}P_{j+2}^{\circ}\right).
\end{multline}
Let us denote the eigenvectors  of the XXZ model as
\begin{equation}
\Psi_{L}(\mathbf{p}_{N})=\sum_{1\leq x_{1}<\dots<x_{N}\leq L}\chi(\mathbf{p}_{N})\left|x_{1},\dots,x_{N}\right\rangle _{L}
\end{equation}
such that
\begin{equation} \label{eq:H1eig}
H_{L}^{(1)}\Psi_{L}^{(1)}(\mathbf{p}_{N})=E(\mathbf{p}_{N})\Psi_{L}^{(1)}(\mathbf{p}_{N}).
\end{equation}
We can construct the eigenvectors of the family $H^{(2)}$ simply by a linear transformation
$\mathcal{F}_{\mathbf{y}_{M}}$ which acts on the basis vectors.

We define for an arbitrary set $x_1,\dots,x_k$
\begin{equation}
  \ket{x_1,\dots,x_k}=\sigma_{x_1}^-\dots\sigma_{x_k}^-\ket{\emptyset}.
\end{equation}
It is important that in writing down the basis states we do not require that the set of coordinates is ordered.

However, we define now the ordered sets $x_1,\dots,x_N$ and $y_1,\dots,y_M$. For a given set $y_1,\dots,y_M$ for which
$y_{i+1}-y_i>1$ and $y_1>2N$ we define the linear transformation as
\begin{multline}
  \mathcal{F}_{\mathbf{y}_{M}}(\left|x_{1},\dots,x_{N}\right\rangle _{L})=\\
 =\left|\tilde{x}_{1},\tilde{x}_{1}+1,
    \dots,\tilde{x}_{N},\tilde{x}_{N}+1,\tilde{y}_{1},\dots,\tilde{y}_{M}\right\rangle _{L+N+M}
\end{multline}
where we defined the shifted coordinates
\begin{align}
\tilde{x}_{k} & =x_{k}+k-1+\sum_{j=1}^{M}\Theta(x_{k}-y_{j}+2N-k+2j-2),\\
\tilde{y}_{j} & =y_{j}-2\sum_{k=1}^{N}\Theta(x_{k}-y_{j}+2N-k+2j-2).
\end{align}
Here $\Theta$ is unit-step function. 
Note that the concatenations of the $x$ and $y$ variables is not ordered in the notations, but the ordering is taken
into account when we actually compute the effective coordinates.

Using this transformation the eigenvectors of the hard rod deformed
Hamiltonian can be found from those of the original one, simply by using the transformation  $\mathcal{F}_{\mathbf{y}_{M}}$.

\begin{widetext}
This is based on the observation
\begin{equation}
\mathcal{F}_{\mathbf{y}_{M}}(H_{L}^{(1)}\left|x_{1},\dots,x_{N}\right\rangle _{L})=H_{L+N+M}^{(2)}\left|\tilde{x}_{1},\tilde{x}_{1}+1,\tilde{x}_{2},\tilde{x}_{2}+1,\dots,\tilde{x}_{N},\tilde{x}_{N}+1,\tilde{y}_{1},\dots,\tilde{y}_{M}\right\rangle _{L+N+M}.
\end{equation}  
Applying the linear transformation $\mathcal{F}_{\mathbf{y}_{M}}$
to (\ref{eq:H1eig}) we obtain that
\begin{equation}
H_{L}^{(2)}\Psi_{L}^{(2)}(\mathbf{p}_{N},\mathbf{y}_{M})=E(\mathbf{p}_{N})\Psi_{L}^{(2)}(\mathbf{p}_{N},\mathbf{y}_{M}),
\end{equation}
where
\begin{equation}
\Psi_{L}^{(2)}(\mathbf{p}_{N},\mathbf{y}_{M})=\mathcal{F}_{\mathbf{y}_{M}}(\Psi_{L-N-M}^{(1)}(\mathbf{p}_{N}))=
\sum_{1\leq x_{1}<\dots<x_{N}\leq L-N-M}\chi(\mathbf{p}_{N})\mathcal{F}_{\mathbf{y}_{M}}(\left|x_{1},\dots,x_{N}\right\rangle _{L-N-M}).
\end{equation}
\end{widetext}
In this way we find all solutions. For fixed $N$ and $M$ we
have $N$ magnon states of the XXZ chain with effective length $L-N-M$.
The Bethe Ansatz solution of the XXZ chain is complete therefore we
obtain 
\begin{equation}
\binom{L-N-M}{N}
\end{equation}
states. We can choose the positions of the DW for a representative
\begin{equation}
\binom{L+1-2N-M}{M}
\end{equation}
ways therefore we found 
\begin{equation}
\sum_{N=0}^{L/2}\sum_{M=0}^{(L-2N+1)/2}\binom{L-N-M}{N}\binom{L+1-2N-M}{M}=2^{L},
\end{equation}
which is the dimension of the Hilbert space.

We can summarize the hard rod deformation as follows: It is a transformation which reshuffles the energy levels and the
eigenstates of the XXZ model into those of the hard rod deformed models with different lengths. Every eigenvector of the
deformed model originates from one of the 
eigenstates of the undeformed model with some smaller volume. This is what we call ``spectral degeneracy across
different models and different volumes''. It is a quantum mechanical generalization of the much simpler phenomenon from
the classical hard rod gas, namely that the only effect of the deformation is that the volume available to the particles
is changed, and the difference depends on the number of the particles within the system. It is remarkable that this
transformation can be formulated on the quantum level, even if we include the particles with length $\ell'=1$ which are
not dynamical on their own.

\subsection{Coordinate Bethe Ansatz -- periodic boundary conditions}

It is also possible to compute the exact wave functions in the periodic case. This will mirror the procedure developed
in \cite{sajat-folded} for the case $\Delta=0$.

We construct excitations above the vacuum state. There are two types of excitations: single particles represented by a
$\bullet$ that are not mobile on their own, and hard rods represented by $\bullet\bullet$ that are dynamical. 
Corresponding to these excitations we introduce local creation operators
\begin{equation}
  \label{Aa}
  A^a_j=
  \begin{cases}
      \sigma^-_j & \text{ if } a=1\\
    \sigma^-_j\sigma^-_{j+1} & \text{ if } a=2.\\
  \end{cases}
\end{equation}
Note that we have automatic exclusions:
\begin{equation}
  A^2_xA^2_{x+1}=A^2_xA^{1}_{x+1}=0.
\end{equation}
Let us consider a state with $N$ hard rods and $M$ single particles; the set of their momenta will be denoted as $\pn$
and $\km$. The wave function can be written down by merging these sets. Therefore we introduce a set of momenta
${\bf q}_{N'}$ and a set of particle types ${\bf a}_{N'}$ with $N'=N+M$.
We assume that there are no coinciding rapidities. 

We find that the wave function computed in \cite{sajat-folded} can be used even in the present case. In real space
representation we have
\begin{equation}
  \label{bondpsi}
  \begin{split}
      \ket{\Psi}=
  \sum_{x_1\le x_2\le \dots \le x_{N'}}
 \sum_{\mathcal{P}\in S_{N'}} e^{i \sum_{j=1}^{N'} q_{\mathcal{P}_j} x_j}  \\
\times \mathop{\prod_{j<k}}_{\mathcal{P}_j>\mathcal{P}_k}
S_{a_j,a_k}(q_{j},q_{k})
\prod_{j=1}^{N'} A^{a_{\mathcal{P}_j}}_{x_j} \ket{\emptyset}.
 \end{split}
\end{equation}
Here $S_{a_j,a_k}(q_{j},q_{k})$ are scattering phase factors, which can be computed from solving the two-particle problems.

We find the following scattering factors for all possible particle pairs:
\begin{equation}
  \begin{split}
    S_{1,1}(q_1,q_2)&=-e^{-i(q_1-q_2)},\\
    S_{2,2}(q_1,q_2)&=e^{-i(q_1-q_2)}S_{XXZ}(q_1,q_2),\\
    S_{1,2}(q_1,q_2)&=e^{-i(q_1-2q_2)},
 \end{split}
\end{equation}
where $S_{XXZ}(q_1,q_2)$ is the scattering factor of the original XXZ model with the same anisotropy $\Delta$ given by
\begin{equation}
  S_{XXZ}(q_1,q_2)=-\frac{e^{i(q_1+q_2)}+1-2\Delta e^{iq_2}}{e^{i(q_1+q_2)}+1-2\Delta e^{iq_1}}.
\end{equation}
This function is usually written in terms of rapidity variables, and then it takes a difference form. However, for our
present purposes it is convenient to keep the momentum variables only.

Apart from the addition of $S_{XXZ}(q_1,q_2)$ these formulas coincide with those given in \cite{sajat-folded}. In fact,
they are exactly reproduced in the limit of $\Delta\to 0$, when $S_{XXZ}(q_1,q_2)\to -1$. 

It was assumed in \eqref{bondpsi} that the original ordering of particle types is $2,2,2,\dots,2,1,1,1,\dots,1$.

The phase factors above are such that for single particles
the occupation of neighbouring sites is forbidden. This ensures that
we do not mistake two single particles with an actual hard rod.

The energy of this state is
\begin{equation}
  E=\sum_{a_j=p} e(p_j).
\end{equation}
The sum runs over the hard rods only.

Let us now also write down the separate Bethe equations for the hard rod and single particle momenta: 
\begin{equation}
  \label{bondBetheeq}
  \begin{split}
    e^{ip_jL}\mathop{\prod_{l=1}^{N}}_{l\ne j} S_{2,2}(p_j,p_l)
    \prod_{m=1}^M S_{2,1}(p_j,k_m)&=1,\\
    e^{ik_lL} \mathop{\prod_{m=1}^M}_{m\ne l} S_{1,1}(k_l,k_m)
    \prod_{j=1}^N S_{1,2}(k_l,p_j)&=1.
  \end{split}
\end{equation}
Substituting the factors we get the simplified Bethe equations
\begin{equation}
  \label{bondBetheeq2}
   \begin{split}
    e^{ip_j(L-N-M)}  \mathop{\prod_{l=1}^{N}}_{l\ne j} S_{XXZ}(p_j,p_l) &=e^{-iP} e^{-2iK},\\
    e^{ik_l(L-2N-M)} &=(-1)^{M-1}e^{-iK}e^{-iP},
  \end{split}
\end{equation}
where we defined
\begin{equation}
  P=\sum_{j=1}^Np_j,\qquad K=\sum_{j=1}^Mk_j.
\end{equation}
We see that the apparent volume is changed for both the hard rods and the immobile single particles, and we also observe
twists to the Bethe equations, which depend on the overall momenta $P$ and $K$. However, apart from this twist
the first equation is completely equivalent to the usual Bethe equations for the hard rod momenta in the modified
volume. This leads to a similar effect as in the open boundary case: some of the eigenvalues of the deformed model will
be given by the spectra of the XXZ chains from smaller volumes. However, the exact correspondence will only work for those
states that have $P+2K=0$, so that there is no apparent twist for the Bethe equations of the hard rod momenta.

\section{Algebraic Construction for $\ell=2$}

\label{sec:aba}

Here we present the algebraic construction of the charges of the hard rod deformed models, by applying the recent
results of \cite{sajat-medium} to these models. We start with a brief review
of the Quantum Inverse Scattering Approach \cite{Korepin-Book}.

\subsection{Standard construction}

The central objects in this framework are the monodromy matrix and the transfer matrix. The monodromy matrix is
constructed as follows. We take an auxiliary space isomorphic to $\complex^d$, where $d$ might or might not be equal to the
dimension of the physical space, which is 2 in our cases. The Lax operator $\La_{a,j}(u)$ is a spectral parameter
dependent operator acting on the tensor product of an auxiliary space (denoted by the index $a$) and a physical space
with site index $j$.  

Then the monodromy matrix for a finite volume $L$ is
\begin{equation}
  M_a(u)=   \La_{a,L}(u)\dots \La_{a,2}(u)\La_{a,1}(u).
\end{equation}
Its trace over the auxiliary space is the transfer matrix:
\begin{equation}
  t(u)=\text{Tr}_a\ M_a(u).
\end{equation}
The transfer matrices with different values of $u$ form a commuting set of operators if the Lax operators satisfy the
exchange relations in the case of two auxiliary spaces $a$ and $b$:
\begin{equation}
  \label{RLL}
  \begin{split}
  R_{b,a}(\nu,\mu) & \mathcal{L}_{b,j}(\nu)  \mathcal{L}_{a,j}(\mu)= \mathcal{L}_{a,j}(\mu) \mathcal{L}_{b,j}(\nu)    R_{b,a}(\nu,\mu).
\end{split}
\end{equation}
Here $R(u,v)$ is the so-called $R$-matrix which satisfies the Yang-Baxter relations
\begin{equation}
  \label{YB}
  \begin{split}
     R_{12}(\lambda_{1},\lambda_2)&R_{13}(\lambda_1,\lambda_3)R_{23}(\lambda_2,\lambda_3)=\\
&  =R_{23}(\lambda_2,\lambda_3) R_{13}(\lambda_1,\lambda_3) R_{12}(\lambda_{1},\lambda_2).
  \end{split}
\end{equation}
In many models the $R$-matrix satisfies the so-called regularity property
\begin{equation}
  R_{ab}(u,u)\sim \Pe_{ab},
\end{equation}
where $\Pe_{ab}$ is the permutation operator acting on the tensor product space. In such cases one can use the
$R$-matrix itself as a Lax operator:
\begin{equation}
\mathcal{L}_{a,j}(\mu)=R_{a,j}(\mu,\xi_0),
\end{equation}
where $\xi_0$ is a fixed parameter of the model. In such cases the YB relation is equivalent to the RLL relation.

The regularity condition ensures that the value of the transfer matrix at the special point $\xi_0$ is
\begin{equation}
  t(\xi_0)=\UU,
\end{equation}
where $\UU$ is the cyclic shift operator on the chain.

A commuting set of local charges is then obtained as
\begin{equation}
  \label{Qalphadef}
  Q_\alpha\quad\sim\quad \left.(\partial_u)^{\alpha-1}\log(t(u)) \right|_{u=\xi_0}.
\end{equation}

Generally different values of $\xi_0$ lead to different integrable models. However, the $R$-matrix is often of
difference form, which means that 
\begin{equation}
  R(u,v)=R(u-v).
\end{equation}
In these cases the parameter $\xi_0$ is irrelevant, and it is conventional to set it to $\xi_0=0$.

Performing the first derivative in \eqref{Qalphadef} we obtain the Hamiltonian
\begin{equation}
  H=Q_2=\sum h_{j,j+1}
\end{equation}
with
\begin{equation}
  h_{j,j+1}=\left.\partial_u \check R_{j,j+1}(u)\right|_{u=0},
\end{equation}
where
\begin{equation}
  \check R(u)=\Pe_{j,j+1} R_{j,j+1}(u).
\end{equation}

\subsection{Modification for 3 site interacting models}

For the three site interacting models we apply a construction which was first published in
\cite{sajat-medium} (with certain elements already appearing in \cite{sajat-cellaut}). This framework allows for the
embedding of the medium range spin chains into the QISM 
method. 

In the case of the three site interacting models we  take two auxiliary spaces $a$ and $b$, a physical space $j$ and a
Lax operator $\La_{a,b,j}(u)$ that acts on the tensor 
product of these three spaces.
Using this Lax operator we define a new transfer matrix acting on $L$ spins as
\begin{equation}
  \label{jot}
  t(u)=\text{Tr}_{a,b}  \left[  \La_{a,b,L}(u)   \dots   \La_{a,b,2}(u)\La_{a,b,1}(u)\right].
\end{equation}
The trace is taken over both auxiliary spaces.

The commutativity of the transfer matrices is proven by finding an $R$-matrix that intertwines the Lax operators in the
so-called RLL relations. Let us label the pairs of auxiliary spaces as $A=ab$ and $B=cd$.
Then the corresponding version of \eqref{RLL} is
\begin{equation}
  \label{RLL2}
  \begin{split}
  R_{B,A}(v,u) & \mathcal{L}_{B,j}(v)  \mathcal{L}_{A,j}(u)= \mathcal{L}_{A,j}(u) \mathcal{L}_{B,j}(v)    R_{B,A}(v,u).
\end{split}
\end{equation}
Here $R_{B,A}(v,u)$ acts on the product of the two paired auxiliary spaces $A$ and $B$, which are themselves 4 dimensional. 
Thus $R$ is a matrix of size $16\times 16$.

For local spin chains with three site interaction
the Lax operator has the initial condition
\begin{equation}
\label{Laxinit}
  \La_{a,b,j}(0)= \Pe_{a,j} \Pe_{b,j},
\end{equation}
which translates into the initial condition $t(0)=\UU^2$ for the transfer matrix. 

We define the conserved charges from the derivatives
\begin{equation}
   \left.(\partial_u)^{\alpha-2}\log(t(u)) \right|_{u=0}.
\end{equation}
Using the initial condition \eqref{Laxinit} we obtain that the Hamiltonian is
\begin{equation}
  H=Q_3=\sum_j h_{j,j+1,j+2}
\end{equation}
where
\begin{equation}
  \label{hLa}
  h_{j,j+1,j+2}=\partial_u\left. \check \La_{j,j+1,j+2}(u)  \right|_{u=0},
\end{equation}
with
\begin{equation}
  \label{Lacsekk}
  \La_{1,2,3}(u)=\Pe_{1,3}\Pe_{2,3} \check \La_{1,2,3}(u).
\end{equation}

In our concrete case the Lax operator is constructed from the corresponding operator of the XXZ model. We found that the
following Lax operator describes the integrable hard rod deformation:
\begin{equation}
  \label{hrdefXXZ1}
      \check \La_{1,2,3}(u) = \check \La^{(XXZ)}_{1,3}(u)P^\bullet_2+  P^\circ_2
\end{equation}
Here $\check \La^{(XXZ)}(u)$ is the Lax operator of the original XXZ spin chain, which is given by
\begin{equation}
  \label{LaXXZ}
  \check \La^{(XXZ)}(u)=
  \begin{pmatrix}
   1 & & & \\
    &\frac{\sinh(\eta)}{ \sinh(u+\eta)} & \frac{\sinh(u)}{ \sinh(u+\eta)}& \\
    &\frac{\sinh(u)}{ \sinh(u+\eta)} & \frac{\sinh(\eta)}{\sinh(u+\eta)}& \\
    & & & 1\\
  \end{pmatrix}.
\end{equation}
Taking the derivative of the Lax operator as described by \eqref{hLa} we obtain the Hamiltonian \eqref{Hhrdef2}.

The $R$-matrix of the model can be found by solving the linear equation \eqref{RLL2}. We obtain the explicit solution as
\begin{widetext}
\begin{equation}
\label{R-hrdXXZ}
R(\lambda,\mu)=
\begin{pmatrix}
E_{11}+E_{44}\rho_1 & E_{21}+E_{43}\rho_2       & E_{31}              & E_{41}\rho_5\\
E_{12}              & E_{22}+E_{44}\rho_6       & E_{32}              & E_{42}\rho_5\\
E_{13}+E_{24}\rho_2 & E_{23}\rho_3              & E_{33}+E_{44}\rho_6 & E_{21}\rho_4+E_{43}\rho_5\\
E_{14}\rho_5        & E_{13}\rho_4+E_{24}\rho_5 & E_{34}\rho_5        & E_{11}\rho_7+(E_{22}+E_{33})\rho_6+E_{44}
\end{pmatrix},
\end{equation}
with the functions
\begin{equation}
\begin{split}
  &\rho_1=\frac{\sinh(\lambda-\mu)\sinh(\mu)}{\sinh(\lambda-\mu+\eta)\sinh(\mu-\eta)},\qquad
  \rho_2=-\frac{\sinh(\lambda-\mu)\sinh(\eta)}{\sinh(\lambda-\mu+\eta)\sinh(\mu-\eta)},\\
&\rho_3=\frac{1}{\sinh(\lambda-\mu+\eta)}\left(\frac{\sinh(\eta)\sinh(\eta+\mu)}{\sinh(\eta+\lambda)}+\frac{\sinh(\lambda-\mu)\sinh(\mu)}{\sinh(\mu-\eta)}\right),\\
&\rho_4=\frac{\sinh(\lambda-\mu)\sinh(\eta)}{\sinh(\lambda-\mu+\eta)\sinh(\lambda+\eta)},\qquad \rho_5=\frac{\sinh(\eta)}{\sinh(\lambda-\mu+\eta)},\\
&\rho_6=\frac{\sinh(\lambda-\mu)}{\sinh(\lambda-\mu+\eta)},\qquad \rho_7=\frac{\sinh(\lambda-\mu)\sinh(\lambda)}{\sinh(\lambda-\mu+\eta)\sinh(\lambda+\eta)}.
\end{split}
\end{equation}
\end{widetext}

It is noteworthy that the $R$-matrix satisfies the factorization condition
\begin{equation}
  \label{fact1}
  \check R_{12,34}(\lambda,0) = \check \La_{2,3,4}(\lambda) \check \La_{1,2,3}(\lambda).
\end{equation}
This condition was derived in \cite{sajat-medium}.

The Lax operator satisfies the inversion relation
\begin{equation}
   \check \La_{j,j+1,j+2}(u) \check \La_{j,j+1,j+2}(-u)=1.
 \end{equation}
As explained in \cite{sajat-medium}, it follows from this relation and the construction of the transfer matrix that the
 operator density of the next charge $Q_5$ is 
 given by formula \eqref{q5relation} announced earlier.

In the special case of $\Delta=1$ we find rational expressions, as expected. These formulas can be derived from the scaling
limit $\lambda \to \eta\lambda$ and $\eta\to 0$. For the Lax operator we find
\begin{equation}
  \label{hrdefXXX}
      \check \La_{1,2,3}(u) = \check \La^{(XXX)}_{1,3}(u)P^\bullet_2+  P^\circ_2
\end{equation}
where now
\begin{equation}
  \label{LaXXX}
  \check \La^{(XXX)}(u)=\frac{1+u\Pe}{1+u}=
  \begin{pmatrix}
   1 & & & \\
    &\frac{1}{1+u} & \frac{u}{1+u}& \\
    &\frac{u}{1+u} & \frac{1}{1+u}& \\
    & & & 1\\
  \end{pmatrix}.
\end{equation}
The resulting Hamiltonian is
\begin{equation}
  h_{1,2,3}=(\Pe_{1,3}-1)P^\bullet_2.
\end{equation}
An alternative expression of the Lax operator \eqref{hrdefXXX} is simply
\begin{equation}
     \check \La_{1,2,3}(u)=1 + \frac{u}{1+u}h_{1,2,3}
\end{equation}
For the $R$-matrix we find in this special case
  \begin{multline}
    \check R_{12,34}(u,v)=1+\frac{u-v}{u-v+1}\Big(h_{234}+h_{123}+
      \\ +\frac{u}{u+1}h_{234}h_{123}+\frac{v}{v-1}h_{123}h_{234}\Big).
\end{multline}
The factorization condition \eqref{fact1} can be checked easily after substituting $v=0$.

\subsection{Chiral charges in the folded XXZ model}

\label{sec:chiral}

The folded XXZ model corresponds to the point $\Delta=0$, and as we discussed above, it also coincides with a
special point of the Bariev model. In our framework this point is interpreted as the hard rod deformation of the XX
model.  In this case  the set
of conserved charges is larger than for finite $\Delta$, and this mirrors the situation known from the standard XX model
\cite{integrability-test}.  Let us therefore discuss the algebraic structure of this model in more detail.

The Lax operator for the canonical charges can be obtained from those presented above by the direct substitution
$\Delta=0$ or equivalently $\eta=i\pi/2$; the formulas were already presented in \cite{sajat-cellaut}. The $R$-matrix
that we find actually coincides with the $U\to 1$ limit of the $R$-matrix of the Bariev model presented in
\cite{bariev-lax-1}, but the limiting procedure is delicate. 

However, as stated above, in this model the set of the conserved charges is actually doubled. 
The canonical charges derived from the transfer matrix above can be written as
\begin{equation}
  Q_\alpha=Q_\alpha^+ +(-1)^{\frac{\alpha+1}{2}} Q_\alpha^-,
\end{equation}
where the two chiral parts $Q_\alpha^+$ and $Q_\alpha^-$ move particles only to the right and left, respectively.
It was already observed in \cite{folded1}, that for this model the chiral charges are separately conserved, similar to
the situation in the original XX model \cite{integrability-test}. Nevertheless 
the algebraic construction behind the chiral charges of the folded XXZ model remained unknown. We now provide this
construction. 

We start with the elementary chiral move of a single hard rod:
\begin{equation}
  M_j=\sigma^+_jP^\bullet_{j+1}\sigma^-_{j+2}.
\end{equation}
These operators satisfy the relations
\begin{equation}
  \label{cell2}
  M_j^2=0, \qquad [M_j,M_k]=0\quad\text{ for }\quad |j-k|>2
\end{equation}
and also 
\begin{equation}
   M_{j}M_{j+1}=0,\qquad M_j M_{j+2}M_{j+1}=0.
\end{equation}
The first chiral charge can be written as
\begin{equation}
  Q_3^+=\sum_j M_j.
\end{equation}
We find that the next chiral charge with range 5 is
\begin{equation}
  \begin{split}
    Q_5^+&=
\sum_j [M_{j+1}+M_{j+2},M_j]=\\
   &= \sum_j \sigma_{j}^+\sigma^+_{j+1}\sigma^-_{j+2}\sigma^-_{j+3}+
    \sigma_j^+ P^\bullet_{j+1}\sigma^z_{j+2}P^\bullet_{j+3}\sigma_{j+4}^-. \\
  \end{split}
\end{equation}
The corresponding space reflected charges are
\begin{equation}
  Q_3^-=(Q_3^+)^\dagger,\qquad Q_5^-=(Q_5^+)^\dagger.
\end{equation}
Direct computation shows that all four charges commute with each other, and the Hamiltonian of the model is
\begin{equation}
  H=Q_3=Q_3^-+Q_3^+.
\end{equation}

Let us now define a formal ``height'' or ``center of mass'' operator as
\begin{equation}
  \dipo=\sum_j \  j  P^\bullet_j.
\end{equation}
All the charges defined above are eigenoperators under the commutator with $\dipo$. Specifically we have
\begin{equation}
  \label{diporel}
  \begin{split}
    [\dipo,Q_3^\pm]&=\pm 2Q_3^\pm\\
    [\dipo,Q_5^\pm]&=\pm 4Q_5^\pm.\\
  \end{split}
\end{equation}
Thus the chiral charges change the center of mass by a well defined amount.

The structure of the first two chiral charges suggests that we can find a Lax operator using our formalism for medium
range chains, and that this operator could be a rational function of the spectral parameter. Indeed we found that the
Lax operator given by \eqref{Lacsekk} and 
\begin{equation}
 \check \La^\pm_{1,2,3}(u)= 1 \pm u \sigma^\pm_1 P^\bullet_2 \sigma^\mp_3
\end{equation}
satisfies the requirements. The chiral transfer matrices defined as
\begin{multline}
  \label{jotch}
  t^s(u)=\text{Tr}_{a,b}  \left[  \La^s_{a,b,L}(u)   \dots   \La^s_{a,b,2}(u)\La^s_{a,b,1}(u)\right], \\
  s=\pm 1
\end{multline}
form commuting families
\begin{equation}
  [t^p(u),t^s(v)]=0,
\end{equation}
where $p=\pm$ and $s=\pm$ independently. The chiral charges are given by their logarithmic derivatives.

The commutativity of the transfer matrices follows from the Yang-Baxter relations. For $s,p=\pm$ we find that 
\begin{equation}
  \label{RLLch1}
  \begin{split}
  R_{B,A}^{sp}(v,u) & \mathcal{L}^s_{B,j}(v)  \mathcal{L}^p_{A,j}(u)= \mathcal{L}^p_{A,j}(u) \mathcal{L}^s_{B,j}(v)    R_{B,A}^{sp}(v,u)
\end{split}
\end{equation}
is satisfied by the following chiral-chiral
\begin{equation}
 \check R^{++}(u,v) = 1 + (u-v)
 \begin{pmatrix}
  0 & E_{4,3} & 0 & u E_{4,1} \\
  0 & 0 & 0 & (E_{3,1}+E_{4,2}) \\
  0 & 0 & 0 & E_{4,3} \\
  0 & 0 & 0 & 0 
 \end{pmatrix}
\end{equation}
and chiral-antichiral $R$-matrices
\begin{widetext}
\begin{equation}
 \check R^{+-}(u,v) =1 + \\
 \begin{pmatrix}
  -uv(E_{1,1}+E_{2,2}+E_{3,3}) & u E_{4,3} & 0 & u^2 E_{4,1} \\
  v E_{3,4} & -uv(E_{1,1}+E_{2,2}) & 0 & u(E_{3,1}+E_{4,2}) \\
  0 & 0 & -uv(E_{1,1}+E_{2,2}+E_{3,3}) & u E_{4,3} \\
  v^2 E_{1,4} & v(E_{1,3}+E_{2,4}) & v E_{3,4} & 0 
 \end{pmatrix}.
\end{equation}
\end{widetext}
The remaining two $R$-matrices read as 
\begin{align}
\check{R}^{--}(u,v) &=\check{R}^{++}(v,u)^T, \\
\check{R}^{-+}(u,v) &=\check{R}^{+-}(v,u)^{-1}.
\end{align}
The Yang-Baxter equations were checked by direct substitution using the program \texttt{Mathematica}.

The structure of the Lax operators imply the following generalization of relations \eqref{diporel}:
\begin{equation}
    [\dipo,Q_\alpha^\pm]=\pm  \frac{\alpha+1}{2} Q_\alpha^\pm.\\
\end{equation}

\section{Hard rod deformation with $\ell=3$}

\label{sec:ell3}

Now the Hamiltonian is given by
\begin{equation}
  H=\sum_j h_{j,j+1,j+2,j+3}
\end{equation}
with the 4-site operator
\begin{multline}
  \label{Hhrdef3}
h_{j,j+1,j+2,j+3}=(\sigma^-_j\sigma^+_{j+3}+\sigma^+_j\sigma^-_{j+3})P^\bullet_{j+1}P^\bullet_{j+2}\\
   -\Delta (P^\circ_j P^\bullet_{j+1}P^\bullet_{j+2}P^\bullet_{j+3}+P^\bullet_j P^\bullet_{j+1}P^\bullet_{j+2}P^\circ_{j+3}).
\end{multline}
We identify $Q_4=H$. The next charge is a 7-site operator, and its density is given by
\begin{equation}
  q_{7}(j)\sim [h(j), h(j+1)+h(j+2)+h(j+3)].
\end{equation}
Based on the earlier result we conjecture that the associated Lax operator takes the form
\begin{equation}
  \label{Lachekk4}
  \La_{1,2,3,4}(u)=\Pe_{1,4}\Pe_{2,4}\Pe_{3,4} \check \La_{1,2,3,4}(u)
\end{equation}
with
\begin{equation}
  \label{hrdefXXZ2}
      \check \La_{1,2,3,4}(u) =
      \check \La_{1,4}(u)P^\bullet_2P^\bullet_3+
    \left(\mathbb{I}_{23}-P^\bullet_2P^\bullet_3\right).
\end{equation}
where $\check \La_{1,4}(u)$ is again the Lax operator of the XXZ chain \eqref{LaXXZ}, but now acting on the spaces 1 and
4. The 
transfer matrix is then constructed using 3 auxiliary spaces as
\begin{equation}
  \label{jot4}
  t(u)=\text{Tr}_{a,b,c}  \left[  \La_{a,b,c,L}(u)   \dots   \La_{a,b,c,2}(u)\La_{a,b,c,1}(u)\right],
\end{equation}
where $1\dots L$ denote the physical spaces.

This model describes the interaction of dynamical hard rods of length $\ell=3$, but it also supports particles of length
1 and 2. These latter particles are not dynamical on their own, but again they do interact with the hard rods. A full
Bethe Ansatz solution can be given analogously to the case of $\ell=2$, but we do not pursue this direction here.

\section{Conclusions}

\label{sec:concl}

In this paper we studied new integrable spin-1/2 chains, which can be considered as hard rod deformations of the
interacting XXZ spin chains. The folded XXZ model \cite{folded1,sajat-folded} is a special member of this family,
corresponding to the hard rod deformation of the free XX chain, this explains its relative simplicity. However, we found
that many of the special features of that model are also displayed by the interacting cases: for example, we observe
exponential degeneracies for excited states, Hilbert space fragmentation, and we also conjecture that persistent
oscillations could be found in certain non-equilibrium problems.

One of the most interesting phenomena is the presence of the exact degeneracies across different deformations and
volumes. This phenomenon has a clear classical counterpart: if we have systems with hard rods, then the available volume
is the total volume minus the sum of the lengths of the rods, and this apparent volume determines the finite volume
properties of the system. We find that the same effect holds in the quantum case, with exact degeneracies in the open
boundary case. It is remarkable that the degeneracies are indeed exact, and that the wave functions of the deformed
models can be obtained by a simple non-local transformation from the original, undeformed ones; this was discussed in
detail in Section \ref{sec:boundary}. While it is not surprising to see a quantum version of the hard rod deformation,
we believe that it is quite remarkable to find these exact correspondences in our truly interacting spin chains with
strictly local Hamiltonians.

We applied the methods of \cite{sajat-medium} to prove the algebraic integrability of the
models. As a by-product we also found the full set of chiral charges for the folded XXZ model, which were not yet found
in \cite{folded1,sajat-folded}. 

There are various open questions. Perhaps the most interesting one is whether hard rod deformations exist for other
integrable spin chains. As we explained in the Introduction, the hard rod deformation is a special generalization of the
famous $T\bar T$ deformation. It is known that other generalizations exist on every integrable chain
\cite{sajat-ttbar,sfondrini-ttbar}, but the question is open for those deformations which involve the momentum, such as
the hard rod or the actual $T\bar T$ deformation. If the hard rod deformation exists for other spin chains, it would be
a general method to construct a new medium range integrable chain starting from a known nearest neighbour model. Somewhat
similar ``model generating'' transformations were studied recently in \cite{clifford1}. The specific form of our Lax
operators suggests that perhaps the hard deformation is more general and not restricted to the
XXZ chains, but this question needs further study. 

Also, it would be useful to get a better understanding of the algebra behind our Lax operator and the
$R$-matrices. Our Lax operators have a very suggestive structure which mirrors the real space representation of the hard
rod deformation. However, we don't have a transparent representation of our $R$-matrices. It would be
interesting to clarify their origin and their interpretation.

\begin{acknowledgments}
  We acknowledge collaboration with Eric Vernier and Yunfeng Jiang on the early staged of this project, and we are
  thankful to Lenart Zadnik for useful discussions.
\end{acknowledgments}


%

\end{document}